\documentclass[
reprint,
 amsmath,amssymb,
 aps,
 pra,
]{revtex4-1}
\usepackage{graphicx}
\usepackage{dcolumn}
\usepackage{bm}
\usepackage{caption}
\usepackage{subcaption}

\begin{document}
\preprint{APS/123-QED}

\title{Quantum mechanical uncertainties and exact transition amplitudes for time dependent quadratic Hamiltonian}

\author{Gal Harari}
 \email{gharari@tx.technion.ac.il}
\author{Yacob Ben-Aryeh}
 \email{phr65yb@physics.technion.ac.il}
\author{Ady Mann}
 \email{ady@physics.technion.ac.il}
\affiliation{Department of Physics, Technion-Israel Institute of Technology, 32000 Haifa, Israel}

\begin{abstract}
In this work we present the simplest generic form of the propagator for the time-dependent quadratic Hamiltonian. We manifest the simplicity of our method by giving explicitly the propagators for a free particle in time-dependent electric field, forced harmonic oscillator and the Paul trap. Exact transition amplitudes and uncertainties are calculated analytically for the Paul trap and harmonic oscillator. The results show that near the instability regions very large quantum mechanical uncertainties are obtained as demonstrated in a special figure. The method is also applied to calculating the trajectory of a classical forced time-dependent harmonic oscillator.
\end{abstract}

\maketitle
\section{introduction}
In recent years experiments with trapped ions have made important contributions to many fields in physics \cite{Leibfried03}. As is well known, squeezing effects related to quantum mechanical uncertainties \cite{Dodonov03}
have been found in such systems. In the present article quantum mechanical uncertainties and exact transition probabilities for a general time dependent quadratic Hamiltonian are calculated, putting special emphasis on those relevant to Paul ion trap \cite{Paul90}. We exhibit in the analysis the interesting effect of large quantum mechanical uncertainties in regions close to the instability region. Although the average values of $x$ and $p$ follow the classical trajectories the uncertainties exhibit significant quantum mechanical behavior. Our analysis has important implications, e.g., the use of mass spectrometry can be applied only in the regions of small quantum mechanical fluctuations~\cite{March97} which can be calculated exactly by our method; a trapped ion can be stored as long as the quantum fluctuations are smaller than the trap dimension. Of course, the time-dependent quadratic Hamiltonian is basic in all fields of physics and our new simplified expression for its exact propagator should be very useful.

In our previous work~\cite{Harari11} the general quadratic time-dependent Hamiltonian was considered
\begin{equation}\label{eq:Hamiltonian}
\hat{H}=a(t)\hat{X}^2+b(t)\left(\hat{X}\hat{P}+\hat{P}\hat{X}\right)+c(t)\hat{P}^2+d(t)\hat{X}+e(t)\hat{P}
\end{equation}
with $a,b,d,e$ arbitrary real functions of time and $c$ a real positive function of time. The following lowering operator (and its Hermitian conjugate) was shown to be invariant (this $linear$ invariant has a long history~\cite{Malkin70,Dodonov89,Dodonov03,Glauber92,Suslov10})
\begin{equation}\label{eq:C}
\hat{C}(t)=\frac{i}{2}\sqrt{\frac{c(0)}{\hbar\nu}}\left(\frac{2bu-\dot{u}}{c}\hat{X}+2u\hat{P}+\frac{ue}{c}-\zeta_{t,1}\right),
\end{equation}
with $u(t)$ the solution of the classical equation of motion without the linear terms
\begin{equation}\label{eq:u}
\ddot u=\frac{\dot c}{c}\dot u+\left(4b^2+2\dot b-2\frac{\dot c}{c}b-4ca\right)u,
\end{equation}
subjected to the initial conditions
\begin{equation}\label{eq:uinitcond}
u(0)=1, \qquad \dot u(0)=i\nu, \qquad \nu > 0.
\end{equation}
and
\begin{equation}
\zeta_{t,1}=\int_{t_{1}}^{t}\frac{u}{c}\left[\dot{e}+e\left(2b-\frac{\dot{c}}{c}\right)-2cd\right]\mathrm{d}t'.
\end{equation}
Using this invariant a complete set of solutions to the time-dependent Schr\"odinger equation (TDSE) was constructed and summed using Mehler's formula~\cite{Merzbacher98,Harari11}. The derived propagator is a time-dependent Gaussian,
\begin{equation}\begin{split}\label{eq:Prop}
&G\left(x_{2},t_{2};x_{1},t_{1}\right)=\langle x_2|U(t_2,t_1)|x_1\rangle\\
&=A(t_{2},t_{1})\exp\left({\sum_{i,j=1}^{2}\alpha_{ij}x_{i}x_{j}+\sum_{i=1}^{2}\gamma_{i}x_{i}}\right).
\end{split}\end{equation}
Here $A$, $\alpha_{ij}$ and $\gamma_i$ depend explicitly on $u(t)$.

The results obtained in our previous work can be further simplified using Abel's identity~\cite{DiPrima00}, yielding the simplest form of the generic propagator. $\alpha_{11}$ and $\alpha_{22}$ take the simple form
\begin{equation}\label{eq:alpha22}
\alpha_{22}(t_2,t_1)=\frac{i\Im\left(\dot{u}_{2}u_{1}^{*}\right)}{4\hbar c_{2}\Im\left(u_{2}u_{1}^{*}\right)}-\frac{ib_{2}}{2\hbar c_{2}}
\end{equation}
\begin{equation}\label{eq:alpha11}
\alpha_{11}(t_{2},t_{1})=\frac{i\Im\left(\dot{u}_{1}u_{2}^{*}\right)}{4\hbar c_{1}\Im\left(u_{2}u_{1}^{*}\right)}+\frac{ib_{1}}{2\hbar c_{1}}.
\end{equation}
$\alpha_{21}$ keeps its simple form
\begin{equation}\label{eq:alpha21}
\alpha_{21}=\alpha_{12}=\frac{-i\nu}{4\hbar c(0)\Im\left(u_{2}u_{1}^{*}\right)}.
\end{equation}
Subscripts of the functions ($a,b,c,d,e,u$) stand for time indices, e.g. $u_2\equiv u(t_2)$.
$\gamma_i$ are also simplified and take the form
\begin{equation}\label{eq:gamma2}
\gamma_2(t_2,t_1)=\frac{1}{2i\hbar}\left(\frac{e_{2}}{c_{2}}+\frac{\Im[u_{1}\zeta_{2,1}^{*}]}{\Im[u_{2}u_{1}^{*}]}\right)
\end{equation}
\begin{equation}\label{eq:gamma1}
\gamma_1(t_2,t_1)=\gamma_2^*(t_1,t_2)=-\frac{1}{2i\hbar}\left(\frac{e_{1}}{c_{1}}+\frac{\Im[u_{2}\zeta_{2,1}^{*}]}{\Im[u_{2}u_{1}^{*}]}\right)
\end{equation}
and
\begin{equation}\label{eq:A}
A(t_2,t_1)=\frac{\exp\frac{i}{4\hbar}\int_{t_{1}}^{t_{2}}\left(\frac{e^{2}}{c}-\frac{c\Im^{2}\left(\zeta_{t,1} u_{1}^{*}\right)}{\Im^{2}\left(u_{2}u_{1}^{*}\right)}\right)\mathrm{d}t}{\sqrt{4\pi i\hbar c(0)\frac{\Im\left(u_{2}u_{1}^{*}\right)}{\nu}}}.
\end{equation}
The various terms can be seen to be $\nu$ independent by writing $u$ as
\begin{equation}
u(t)=u_R(t)+i\nu u_I(t)
\end{equation}
with $u_R$ and $u_I$ the real and imaginary parts of the solution to Eq.\eqref{eq:u}, i.e.
\begin{equation}\label{eq:EvenOddInit}
u_R(0)=1,\quad\dot{u}_R(0)=0,\quad u_I(0)=0,\quad\dot{u}_I(0)=1.
\end{equation}
Therefore the imaginary parts are proportional to $\nu$, which cancels in the quotients. As an example for the simplicity of our method the case of the forced simple harmonic oscillator (HO) is directly obtained without the need for path integration~\cite{Feynman2010,Schulman2012} or canonical transformations with,
\begin{equation}\begin{split}\label{eq:PropForcedHO}
&A=\sqrt{\frac{m\omega}{2\pi i\hbar\sin\omega\Delta t}}\times\\
&\exp\left[\frac{i\int_{t_{1}}^{t_{2}}\left(\int_{t_{1}}^{t''}d\left(t'\right)\sin{\omega( t_1-t')}\mathrm{d}t'\right)^2\mathrm{d}t''}{\hbar\sin^{2}\omega\Delta t}\right]\\
&\alpha_{22}=\alpha_{11}=\frac{im\omega\cos\omega\Delta t}{2\hbar\sin\omega\Delta t},\quad \alpha_{21}=-\frac{im\omega}{2\hbar\sin\omega\Delta t},\\
&\gamma_1= \exp\left[{\frac{i\int_{t_{1}}^{t_{2}}d\left(t\right)\sin\omega\left(t-t_{2}\right)\mathrm{d}t}{\hbar\sin\omega\Delta t}}\right],\\
&\gamma_2=\exp\left[\frac{i\int_{t_{1}}^{t_{2}}d\left(t\right)\sin\omega\left(t_{1}-t\right)\mathrm{d}t}{\hbar\sin\omega\Delta t}\right],
\end{split}\end{equation}
with $\Delta t=t_2-t_1$.

Our method is also applicable to cases where the dynamics is not unitary (though still quadratic), such as the case of damping. In such cases the linear invariant so defined is still applicable and so are its eigenfunctions and their sum which yield the (non unitary) propagator. This is demonstrated in the next section for the classical case and can also be applied to the quantum case.
\section{applications}
\subsection{The classical case}
In the classical case (with arbitrary friction term, not necessarily derived from Hamilton equations) we have
\begin{equation}\label{eq:EQM_classical}
\ddot{x}+\alpha(t)\dot{x}+\beta(t)u(x)=f(t).
\end{equation}
Defining $u$ as the solution of the homogeneous equation we write the classical invariant (analogous to Eq.~\eqref{eq:C})
\begin{equation}\label{eq:C_classical}
C(t)=\left(\dot{u}x-\dot{x}u\right)e^{\int_{0}^{t}\alpha(t')\mathrm{d}t'}+\tilde{\zeta}\left(t\right),
\end{equation}
with
\begin{equation}
\tilde{\zeta}\left(t\right)=\int_{0}^{t}u(\tau)f(\tau)e^{\int_{0}^{\tau}\alpha(t')\mathrm{d}t'}\mathrm{d}\tau.
\end{equation}
The classical trajectory is
\begin{equation}\label{eq:x_classical}
x(t)=x_{0}\Re\left(u\right)- v_{0}\frac{\Im\left(u\right)}{\nu}+\frac{1}{\nu}\Im \left( u\tilde{\zeta}\right)
\end{equation}
Thus, the forced motion is obtained by solving the homogenous equation for $u$. In particular, the well known result~\cite{Landau} for the forced harmonic oscillator ($a=\frac{1}{2}m\omega^2,$ $b=0,$ $c=\frac{1}{2m}$) may be achieved directly
\begin{equation}
x(t)=\cos\left(\omega t\right)x_{0}+\frac{\sin\left(\omega t\right)}{\omega}v_{0}+\frac{\int_{0}^{t}\sin\omega\left(t-t'\right)f(t')\mathrm{d}t'}{m\omega}.
\end{equation}

A Paul trap utilizes a combination of DC and AC fields in order to trap an ion (with charge $Z$) in a stable
trajectory. Its (classical) Hamiltonian is quadratic with
\begin{equation}
a=Z\frac{U-V\cos\omega_{rf} t}{2r_{0}^{2}},\quad c=\frac{1}{2m},\quad b=d=e=0;
\end{equation}
 $U$ and $V$ are the DC and AC amplitudes, respectively and $r_0$ is roughly the trap size~\cite{Ghosh96}. The equation of motion in a dimensionless form is the well known Mathieu equation
\begin{equation}
x_{\xi\xi}+\left(\alpha-2q\cos2\xi\right)x=0,\label{eq:Mathieu}
\end{equation}
with $\xi=\frac{\omega_{rf}}{2}t,\quad\alpha=\frac{8ZU}{mr_{0}^{2}},\quad q=\frac{4ZV}{mr_{0}^{2}}$.
Therefore we get for $u$ (see Eq.\eqref{eq:u} )
\begin{equation}\label{eq:uMathieuCl}
u=\text{MC}\left(\alpha,q,\frac{\omega_{rf}}{2}t\right)+i2\frac{\nu}{\omega_{rf}}\text{MS}\left(\alpha,q,\frac{\omega_{rf}}{2}t\right)
\end{equation}
where MC and MS are the even and odd solutions of Eq.\eqref{eq:Mathieu} respectively (satisfying Eq.\eqref{eq:EvenOddInit}). The classical trajectory (including a force term) is
\begin{equation}\begin{split}\label{eq:MathieuCl}
x(t)=&x_{0}\text{MC}\left(\alpha,q,\frac{\omega_{rf}}{2}t\right)-\frac{2v_{0}}{\omega_{rf}}\text{MS}\left(\alpha,q,\frac{\omega_{rf}}{2}t\right)\\
&+\frac{1}{m\nu}\Im\left(u\int_{0}^{t}u^{*}(t')f(t')\mathrm{d}t'\right).
\end{split}\end{equation}
As $\alpha$ and $q$ get closer to the instability region the solution becomes exceedingly large and unstable relative to the initial conditions since $u$ approaches an exponentially divergent solution.
\subsection{Uncertainty of an arbitrary Gaussian}
Given a normalized Gaussian
\begin{equation}\label{eq:GenGauss}
\psi=\left(\frac{2a_{R}}{\pi}\right)^{\frac{1}{4}}\exp\left(-ax^{2}+bx-\frac{b_{R}^{2}}{4a_{R}}\right)
\end{equation}
with $a,b\in\mathbb{C}$ (not to be confused with the functions in the Hamiltonian), $a_{R}=\Re(a)>0,$ $b_R=\Re(b)$, the variances are
\begin{equation}\begin{split}\label{eq:VarXP}
\Delta x=\frac{1}{2}\sqrt{\frac{1}{a_{R}}},\, \Delta p=\hbar\sqrt{\frac{a_{I}^{2}+a_{R}^{2}}{a_{R}}},\, \Delta x\Delta p=\frac{\hbar}{2}\sqrt{1+\frac{a_{I}^{2}}{a_{R}^{2}}}.
\end{split}\end{equation}

Now, given an initial Gaussian with variances $(\Delta x)_0$, $(\Delta p)_0$ one can calculate the variances at any time $t$ since the propagator is also a Gaussian. The resulting $a$ (Eq.\ref{eq:GenGauss}) is
\begin{equation}\label{eq:adxdp}
a(t)=\frac{4\Delta x_{0}^{2}\alpha_{21}^{2}}{4\Delta x_{0}^{2}\alpha_{11}-i\sqrt{4\Delta p_{0}^{2}\Delta x_{0}^{2}/\hbar^{2}-1}-1}-\alpha_{22}.
\end{equation}
The variance of $x$ is particularly simple when the initial Gaussian has minimum uncertainty ($\Delta x_0 \Delta p_0 =\frac{\hbar}{2}$)
\begin{equation}
\frac{\Delta x}{\left(\Delta x\right)_{0}}=\sqrt{\left(\Re(u)-2b_{0}\frac{\Im(u)}{\nu}\right)^{2}+\left(\frac{\hbar c_{0}}{(\Delta x)_0^2}\frac{\Im(u)}{\nu}\right)^{2}}.
\end{equation}
It is seen that in general the smaller the initial uncertainty in position the lesser is the particle confined, as expected from general uncertainty considerations.
\subsection{The Paul trap}
\begin{figure}[!ht]
	\centering
        \begin{subfigure}[b]{0.2\textwidth}
                \centering
                \includegraphics[width=\textwidth]{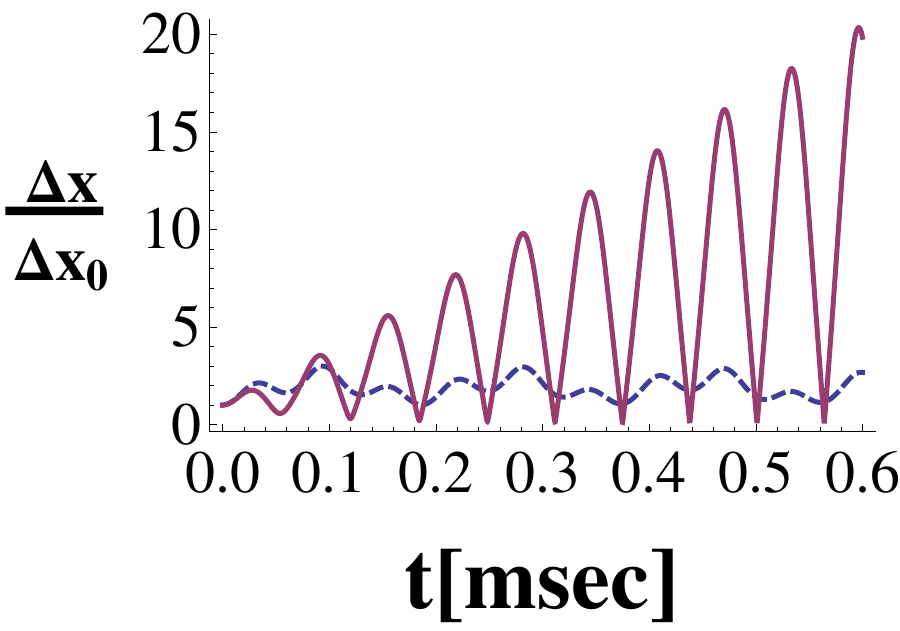}
                \caption{$\Delta x$}
                \label{fig:Dx}
        \end{subfigure}
        \begin{subfigure}[b]{0.2\textwidth}
                \centering
                \includegraphics[width=\textwidth]{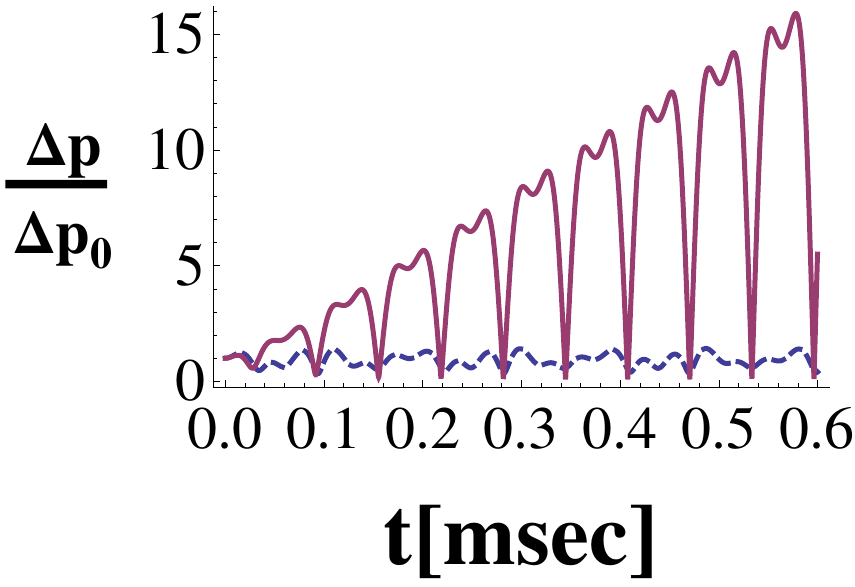}
                \caption{$\Delta p$}
                \label{fig:Dp}
        \end{subfigure}\\
        \begin{subfigure}[b]{0.3\textwidth}
                \centering
                \includegraphics[width=\textwidth]{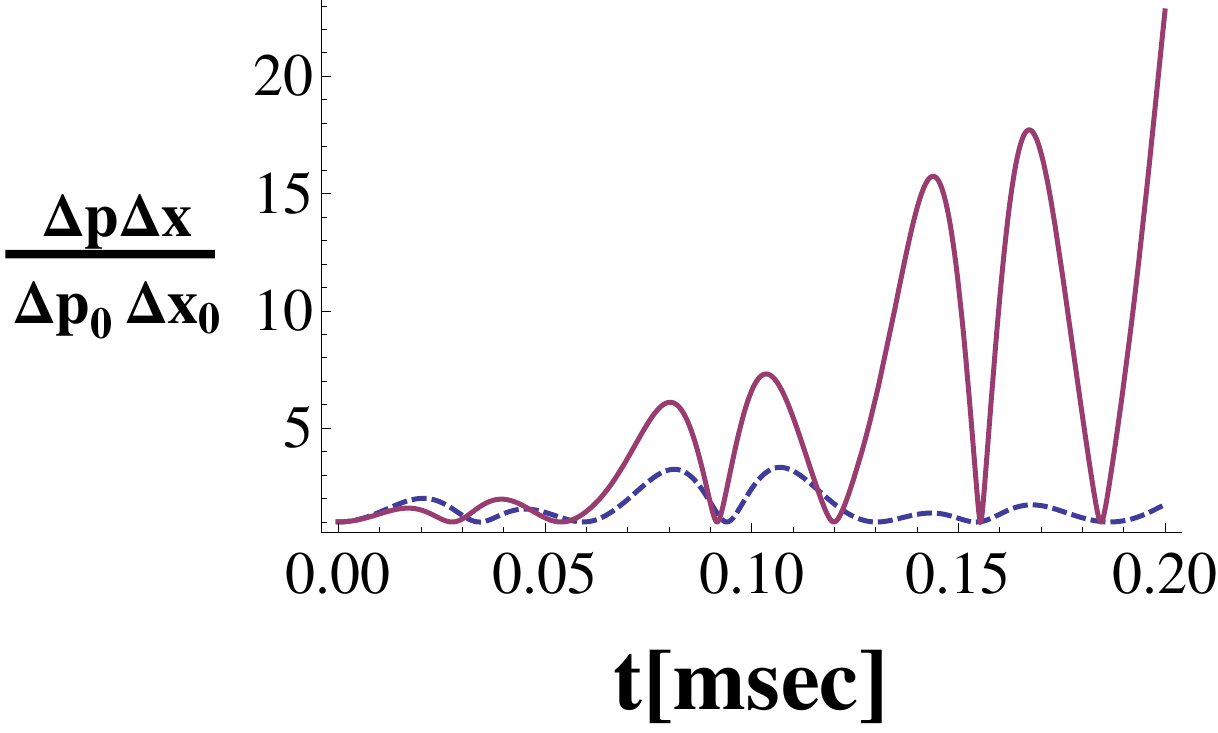}
                \caption{$\Delta x\Delta p$}
                \label{fig:DxDp}
        \end{subfigure}
        	\caption{Relative variances as functions of time for a Gaussian with minimal initial uncertainties and $\Delta x_0=0.1mm$. For trap parameters deep in the stability region (dashed blue) $a=0.005,q=0.46$ and close to the instability region (solid purple) $a=0.37,q=0.5895$. The difference between the two configurations is quite evident; in the latter case the ion is hardly localized in the trap}\label{fig:Variances}
\end{figure}

Having found $u$ (Eq. \ref{eq:uMathieuCl}) the explicit propagator of the forced ion in a Paul trap (where $d(t)$ is the force term) is (choosing $t_{1}=0$ and omitting $\alpha,q$ from the argument of $u$ for simplicity)
\begin{equation}\begin{split}\label{eq:MathieuProp}
&G_{trap}\left(x_{2},t_{2};x_{1},0\right)=\\
&\left(\frac{m\omega_{rf}e^{\frac{-i}{m\hbar}\int_{0}^{t}\frac{\left(\int_{0}^{\tau}\left[\text{MS}\left(\frac{\omega}{2}t'\right)\right]d(t')\mathrm{d}t'\right)^{2}}{\text{MS}^{2}\left(\frac{\omega}{2}t_{2}\right)}\mathrm{d}\tau}}{4\pi i\hbar\text{MS}\left(\frac{\omega_{rf}}{2}t\right)}\right)^\frac{1}{2}\\
&\times\exp\left[\frac{\dot{\text{MS}}\left(\frac{\omega_{rf}}{2}t\right)x_2^{2}-2x_2x_1+\text{MC}\left(\frac{\omega_{rf}}{2}t\right)x_1^{2}}{\frac{4\hbar}{im\omega_{rf}}\text{MS}\left(\frac{\omega_{rf}}{2}t\right)}\right]\\
&\times e^{\frac{i}{\hbar}\left(\frac{\text{MC}\left(\frac{\omega}{2}t\right)}{\text{MS}\left(\frac{\omega}{2}t\right)}\int_{0}^{t}\left[\text{MS}\left(\frac{\omega}{2}\tau\right)\right]d(\tau)\mathrm{d}\tau-\int_{0}^{t}\left[\text{MC}\left(\frac{\omega}{2}\tau\right)\right]d(\tau)\mathrm{d}\tau\right)x_1}\\
&\times e^{\frac{-i\left(\int_{0}^{t}\left[\text{MS}\left(\frac{\omega}{2}\tau\right)\right]d(\tau)\mathrm{d}\tau\right)}{\hbar\text{MS}\left(\frac{\omega}{2}\tau\right)}x_2}.
\end{split}\end{equation}
We can now calculate explicitly the variances as functions of time and trap parameters ($\alpha$, $q$). Uncertainties for typical trap parameters (see e.g.~\cite{Ghosh96,alheit98}) deep inside the stability region and parameters close to the instability region are shown in figure~\ref{fig:Variances}. The particle is less confined in the latter case. The initial uncertainty is taken as $0.1mm$, one order of magnitude less than the trap dimension.
\subsection{Transition amplitudes}
\begin{figure}[ht]
	\centering
    \includegraphics[width=0.45\textwidth]{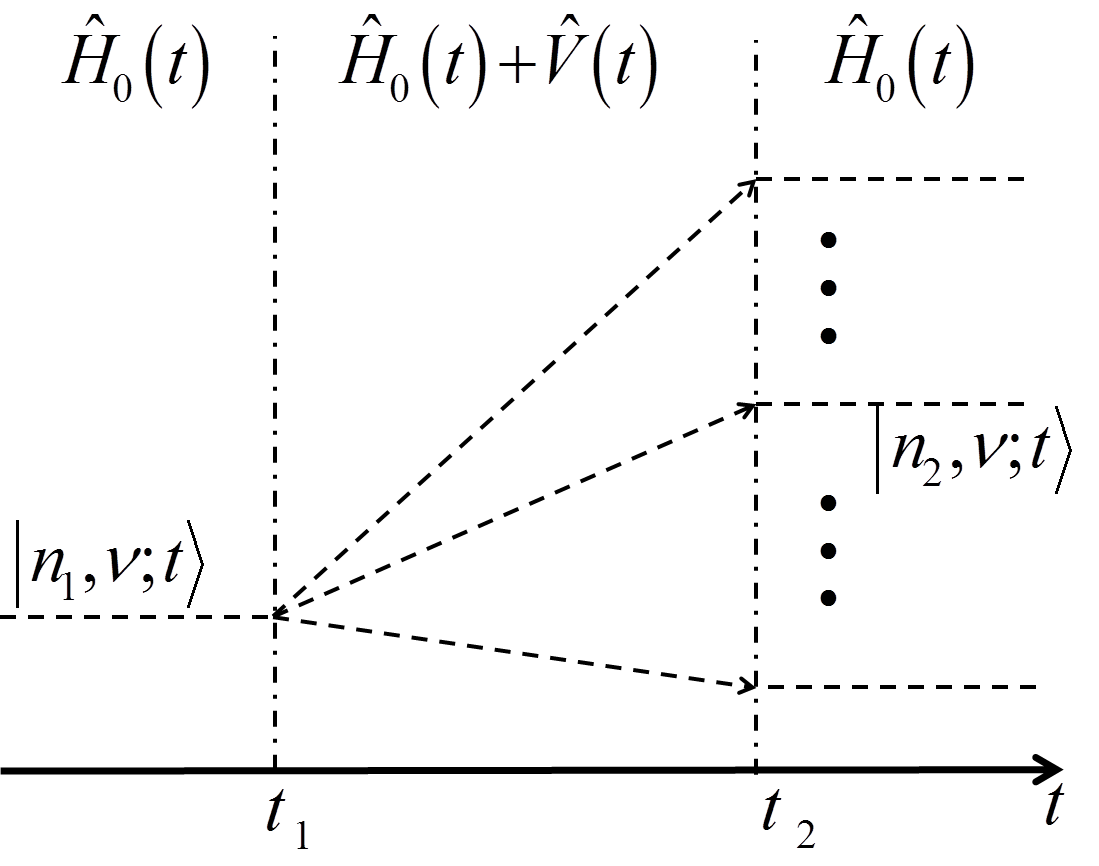}
    \caption{Schematic diagram of the system dynamics. At times $t<t_1$ and $t>t_2$ the system evolves under the quadratic terms only, $\hat{H}_0$. At times $t_1<t<t_2$ the external forces (linear terms) are added, causing transition from the initial state $|n_1;t<t_1\rangle$ to a superposition of states $\sum_{n_2} T_{n_2,n_1}\psi_{n_2,\nu}(t>t_2)$. The transition probability is calculated exactly between any two times.}\label{fig:Tmn}
\end{figure}

In certain cases the solutions of the TDSE (with no applied external force) bear special physical meaning; these $\nu$ dependent wave functions were found in our previous work~\cite{Harari11}
 \begin{equation}\begin{split}\label{psi_n}
&\psi_{n,\nu}(x,t)=\frac{1}{\sqrt{n!}}\left(\frac{\nu}{2\pi\hbar c(0)u(t)^2}\right)^{\frac{1}{4}}\left(\frac{u^*(t)}{2u(t)}\right)^{\frac{n}{2}}\times\\
&\exp\left(\frac{2b(t)u(t)-\dot u(t)}{4i\hbar c(t)u(t)}x^2\right)H_n\left(\sqrt{\frac{\nu}{2\hbar c(0)|u(t)|^2}}x\right).
\end{split}\end{equation}
 For the simple HO choosing $\nu=\omega$ one gets the stationary solutions (times the required time dependent phase) and for the Paul trap one gets the ($\nu$ dependent) quasi-energy states~\cite{Brown91,Harari11}. In general these states evolve in a complicated way, even without force, however they constitute an orthonormal basis at any given instant. Using our method we can calculate the exact transition amplitudes between such states in arbitrary times (not necessarily short or long) caused by an arbitrary external force (see figure~\ref{fig:Tmn} for a schematic scenario). This is of particular interest in cooling atoms where the $|0,t\rangle$ state has the least average energy\cite{Leibfried03}.
The transition amplitude is the projection of the exactly evolved states (including the external force) on states that evolved with no external force, $|n_2,\nu,t\rangle$:

\begin{equation}\label{eq:Tnn}
T_{n_2,n_1}=\langle n_2,t_{2},\nu|U(t_{2},t_{1})|n_1,t_{1},\nu\rangle
\end{equation}
with $U(t_{2},t_{1})$ the time evolution operator including the linear terms. The integration is carried out using the generating function of Hermite polynomials (thanks to the fact that our propagator is a Gaussian) which results in an explicit result

\begin{equation}\begin{split}\label{eq:Tn1n2}
&T_{n_2,n_1}=\frac{\exp\left[\frac{1}{4\hbar}\int_{t_{1}}^{t_{2}}\left(i\frac{e^{2}\left(t\right)}{c\left(t\right)}-\frac{c_{0}}{\nu}\dot{\zeta}_{t,1}^{*}\zeta_{t,1}\right)\mathrm{d}t\right]}{\sqrt{2^{n_{1}+n_{2}}n_{1}!n_{2}!}}\\
&\times\exp\left[i\left(n_{2}\varphi_{2}-n_{1}\varphi_{1}\right)\right]\partial_{r}^{n_{1}}\partial_{s}^{n_{2}}\exp\left[2e^{i\left(\varphi_{1}-\varphi_{2}\right)}rs\right]\times\\
&\exp\left[i\left(\frac{c_{0}}{2\hbar\nu}\right)^{\frac{1}{2}}\left(e^{i\varphi_{1}}\zeta_{2,1}^{*}r-e^{-i\varphi_{2}}\zeta_{2,1}s\right)\right]\big|_{\{r,s\}=0}
\end{split}\end{equation}
With $\varphi_1=\arg(u_1)$, $\varphi_2=\arg(u_2)$. The double derivative may be calculated explicitly using the relation

\begin{equation}\begin{split}
&\partial_{r}^{n_{1}}\partial_{s}^{n_{2}}\left[\exp\left(xs+yr\right)\exp\left(zrs\right)\right]\big|_{\left\{ r,s\right\} =0}=\\
&\sum_{l=0}^{\min\left(n_{1},n_{2}\right)}\frac{n_{1}!n_{2}!}{l!\left(n_{1}-l\right)!\left(n_{2}-l\right)!}y^{n_{1}-l}x^{n_{2}-l}z^{l}
\end{split}\end{equation}
For the case of the simple harmonic oscillator (with $e(t)=0$, see Eq.\eqref{eq:PropForcedHO}) we get
\begin{equation}\begin{split}\label{eq:TnnHO}
&T_{n_2,n_1}=\frac{\exp\left[-\frac{1}{2m\omega\hbar}\int_{t_{1}}^{t_{2}}d\left(t\right)\int_{t_{1}}^{t}e^{i\omega\left(\tau-t\right)}d\left(\tau\right)\mathrm{d}\tau\mathrm{d}t\right]}{\sqrt{2^{n_{1}+n_{2}}n_{1}!n_{2}!}}\\
&e^{i\omega\left(n_{2}t_{2}-n_{1}t_{1}\right)}\partial_{r}^{n_{1}}\partial_{s}^{n_{2}}\times\\
&e^{\left\{ g_{rs}\left(t_{2},t_{1}\right)rs+\left[g_{r}\left(t_{2},t_{1},t_{1}\right)r+g_{r}^{*}\left(t_{2},t_{1},t_{2}\right)s\right]\right\} }\big|_{\{r,s\}=0}.
\end{split}\end{equation}
With
\begin{equation}\begin{split}
&g_{rs}\left(t_{2},t_{1}\right)=2e^{-i\omega\left(t_2-t_1\right)},\\
&g_{r}\left(t_{2},t_{1},t\right)=-i\sqrt{\frac{1}{m\hbar\omega}}\int_{t_{1}}^{t_{2}}e^{i\omega\left(t-\tau\right)}d\left(\tau\right)\mathrm{d}\tau.
\end{split}\end{equation}

For the case of the Paul trap (see Eq.~\ref{eq:uMathieuCl}, where we chose for simplicity $\nu=\frac{\omega_{rf}}{2}$) the transition amplitude can be expressed using the functions (Mathieu Tangent and Mathieu Exponent, for obvious reasons)
\begin{equation}\begin{split}
&\text{MTAN}\left(\alpha,q,\frac{\omega_{rf}}{2}t\right)\equiv\frac{\text{MS}\left(\alpha,q,\frac{\omega_{rf}}{2}t\right)}{\text{MC}\left(\alpha,q,\frac{\omega_{rf}}{2}t\right)},\\
&\text{MExpI}\left(\alpha,q,t\right)\equiv\text{MC}\left(\alpha,q,t\right)+i\text{MS}\left(\alpha,q,t\right).
\end{split}\end{equation}
So the phase and $\zeta_{t,1}$ are
\begin{equation}\begin{split}\label{eq:Tn1n2Paul}
&\varphi\left(t\right)=\text{MTAN}^{-1}\left(\alpha,q,\frac{\omega_{rf}}{2}t\right),\\
&\zeta_{t,1}=-2\int_{t_{1}}^{t}\text{MExpI}\left(\alpha,q,\frac{\omega_{rf}}{2}\tau\right)d\left(\tau\right)\mathrm{d}\tau.
\end{split}\end{equation}
Hence the transition amplitude is given explicitly using Eq.~\eqref{eq:Tn1n2} and the above definitions. It is important to emphasize that the above transitions amplitudes are exact and valid for short and long times (see Brown~\cite{Brown91} for perturbative treatment).

\section{Summary}
A new important effect in the Paul trap has been shown where very large quantum uncertainties are obtained in the region near the instability borders. This is very important since a trapped ion can be stored as long as the quantum fluctuations are smaller than the trap dimension. A simplified expression for the general propagator for time dependent quadratic Hamiltonian has been presented in Eq.~\eqref{eq:Prop}; the different parameters to be used in this equation are given in Eqs.(\ref{eq:alpha22}-\ref{eq:A}). The classical trajectory (including a force term) for the Paul ion-trap has been presented in Eq.~\eqref{eq:MathieuCl}. Uncertainties for a general normalized Gaussian are presented in Eq.~\eqref{eq:VarXP}. Propagating an initial Gaussian with variances $\Delta x_0,\Delta p_0$ we calculated the time-dependent variances using Eq.\eqref{eq:adxdp}. The propagator for the Paul trap was presented in Eq.~\eqref{eq:MathieuProp} and used for calculating variances stemming from an initial Gaussian with minimum uncertainties. We demonstrated the increase of the quantum mechanical uncertainties near the edge of the stability region in Fig.~\eqref{fig:Variances}. In conclusion we found that both classical and quantum mechanical fluctuations occur near the edge of the stability region. Such fluctuations have important implications from the theoretical point of view related to phase transition~\cite{Diedrich87}. The results have also important implications to Paul mass-spectrometry as the measurement in this field can be accurate only in regions with small quantum mechanical fluctuations which can be calculated exactly by our method~\cite{March97}.

Simple expressions for the exact transition amplitudes for the forced HO are presented in Eq.\eqref{eq:TnnHO} and for the forced ion in a Paul trap in Eq.\eqref{eq:Tn1n2Paul}. The transition amplitudes are of special interest, e.g., in quantum optics.

\bibliographystyle{unsrt}

\begin{thebibliography}{10}

\bibitem{Leibfried03}
D.~Leibfried, R.~Blatt, C.~Monroe, and D.~Wineland.
\newblock Quantum dynamics of single trapped ions.
\newblock {\em Rev. Mod. Phys.}, 75:281--324, 2003.

\bibitem{Dodonov03}
V.~V. Dodonov and V.~I. Man'ko.
\newblock {\em Theory of nonclassical states of light}.
\newblock CRC Press, London, 2003.
\newblock and references therein.

\bibitem{Paul90}
Wolfgang Paul.
\newblock Electromagnetic traps for charged and neutral particles.
\newblock {\em Rev. Mod. Phys.}, 62:531--540, Jul 1990.

\bibitem{March97}
R.E. March.
\newblock An introduction to quadrupole ion trap mass spectrometry.
\newblock {\em J. Mass Spectrom.}, 32(4):351--369, 1997.

\bibitem{Harari11}
G.~Harari, Y.~Ben-Aryeh, and A.~Mann.
\newblock Propagator for the general time-dependent harmonic oscillator with
  application to an ion trap.
\newblock {\em Phys. Rev. A}, 84:062104, Dec 2011.

\bibitem{Malkin70}
I.~A. Malkin, V.~I. Man'ko, and D.~A. Trifonov.
\newblock Coherent states and transition probabilities in a time-dependent
  electromagnetic field.
\newblock {\em Physical Review D}, 2(8):1371, 1970.

\bibitem{Dodonov89}
V.~V. Dodonov and V.~I. Man'ko.
\newblock {\em Invariants and the evolution of nonstationary quantum systems}.
\newblock Nova science publishers, Commack, New York, 1989.
\newblock and references therein.

\bibitem{Glauber92}
R.J. Glauber.
\newblock Laser manipulation of atoms and ions.
\newblock In {\em Proceedings of the Internationl School of Physics "Enrico
  Fermi" Course 118}, page 643. North-Holland, Amsterdam, 1992.

\bibitem{Suslov10}
Sergei~K Suslov.
\newblock Dynamical invariants for variable quadratic hamiltonians.
\newblock {\em Phys. Scr.}, 81:1--11, 2010.

\bibitem{Merzbacher98}
E.~Merzbacher.
\newblock {\em Quantum mechanics}.
\newblock Wiley, New York, third edition, 1998.

\bibitem{DiPrima00}
William~E. Boyce and Richard~C. DiPrima.
\newblock {\em Elementary Differential Equations}.
\newblock Wiley, New York, third edition, 2000.

\bibitem{Feynman2010}
R.P. Feynman, A.R. Hibbs, and D.F. Styer.
\newblock {\em Quantum Mechanics and Path Integrals}.
\newblock Dover books on physics. Dover Publications, 2010.

\bibitem{Schulman2012}
L.S. Schulman.
\newblock {\em Techniques and Applications of Path Integration}.
\newblock Dover Books on Physics. Dover Publications, 2012.

\bibitem{Landau}
L.~D. Landau and E.~M. Lifshitz.
\newblock {\em Mechanics}.
\newblock Pergamon Press, Oxford, third edition, 1976.

\bibitem{Ghosh96}
P.~K. Ghosh.
\newblock {\em Ion Traps}.
\newblock Oxford University Press, New York, 1996.

\bibitem{alheit98}
R.~Alheit, T.~Gudjons, S.~Kleineidam, and G.~Werth.
\newblock Some observations on higher-order non-linear resonances in a {P}aul
  trap.
\newblock {\em Rapid Commun. Mass Spectrom.}, 10(5):583--590, 1998.

\bibitem{Brown91}
Lowell~S. Brown.
\newblock Quantum motion in a {P}aul trap.
\newblock {\em Phys. Rev. Lett.}, 66:527--529, 1991.

\bibitem{Diedrich87}
F.~Diedrich, E.~Peik, J.~M. Chen, W.~Quint, and H.~Walther.
\newblock Observation of a phase transition of stored laser-cooled ions.
\newblock {\em Phys. Rev. Lett.}, 59:2931--2934, Dec 1987.

\end{thebibliography}

\end{document}